\documentclass[review]{elsarticle}

\usepackage{makecell}
\usepackage{graphicx,float,wrapfig,subfigure}
\usepackage{booktabs}
\usepackage{lineno,hyperref}
\usepackage{multirow}
\modulolinenumbers[5]

\journal{Journal of \LaTeX\ Templates}

\begin{document}

\begin{frontmatter}

\title{Nuclear reaction measurements of 80.5 MeV/u $\rm ^{12}C$ beam bombarding on C, W, Cu, Au, Pb targets}

\author[mymainaddress,mysecondaryaddress]{Zhichao Gao}

\author[mymainaddress,mysecondaryaddress]{Xueying Zhang}
\author[mymainaddress]{Yongqin Ju}
\author[mymainaddress]{Liang Chen}
\author[mymainaddress]{Honglin Ge\corref{mycorrespondingauthor}}\cortext[mycorrespondingauthor]{Corresponding author}
\ead{hlge@impcas.ac.cn}
\author[mymainaddress]{Yanbin Zhang}
\author[mymainaddress]{Fei Ma}
\author[mymainaddress]{Hongbin Zhang}
\author[mymainaddress]{Guozhu Shi}
\author[mymainaddress,mysecondaryaddress]{Zhiqiang Chen}
\author[mymainaddress,mysecondaryaddress]{Rui Han}
\author[mymainaddress]{Guoyu Tian}
\author[mymainaddress]{Fudong Shi}
\author[mymainaddress,mysecondaryaddress]{Bingyan Liu}
\author[mymainaddress,mysecondaryaddress]{Xin Zhang}

\address[mymainaddress]{Institute of Modern Physics, Chinese Academy of Science,
Lanzhou 730000, China}
\address[mysecondaryaddress]{School of Nuclear Science and Technology, University of Chinese Academy of Sciences, Beijing 100049,
China}

\begin{abstract}
To get the energy spectrum distribution and cross-sections of emitted light charged particles and explore the nuclear reaction, a experiment of 80.5 MeV/u $\rm ^{12}C$ beam bombarding on C, W, Cu, Au, Pb targets has been carried out at Institute of Modern Physics, Chinese Academy of Science. $30^{\circ}$, $60^{\circ}$ and $120^{\circ}$ relative to the incident beam have been detected using three sets of telescope detectors. The method of $\Delta E$-$\Delta E$-$E$ was used for particle identification. The results indicate that there is a tendency that heavier targets have larger double differential cross-sections and the emitted fragments are more likely to go forward. Besides, the decrease of cross-sections of fragments producing with the increasing emitted angles may follow some kind of pattern.
\end{abstract}

\begin{keyword}
\texttt nuclear reaction\sep heavy target \sep energy spectrum\sep cross-section
\end{keyword}

\end{frontmatter}

\linenumbers

\section{Introduction}
Heavy ion nuclear reactions refers to the nuclear reactions caused by ions heavier than $\alpha$ particles. With the development and improvement of heavy ion accelerator technology, recently, people have been able to accelerate almost all the stable nuclides existing in the nature to the energy sufficient to cause nuclear reaction. The possible reaction systems have been greatly expanded, and abundant experimental phenomena not found in light ion nuclear reactions have been observed. Nowadays heavy ion nuclear reaction has become the main field of nuclear physics research~.
~The heavy ion  nuclear reaction offers an opportunity to study phase transition in nclear matter~\cite{POCHODZALLA1997443}, the nuclear quation of the state~\cite{GIULIANI2014116} and the properties of high temperature and high density nuclear matter~\cite{DITORO2007267}. As carbon is an important componet of cosmic rays~\cite{elemental}, experiments of carbon beams bombarding on targets are also used to study the radiation risks of people on space missions. 

For the purpose of improving the dose deposition accuracy of hadron therapy, many countries have carried out experiments of carbon beams of different energies bombarding on varieties of targets. 
Measurements of total and partial cross-sections of fregments for incident $\rm^{12}C$ beams in water and polycarbonate at energies ranging from 200 to 400 MeV/u were made by Japan scientists (T. TOSHITO et al.) in 2004~\cite{05}.
In 2008, to extend these data to lower energy, France scientists performed an experiment of a 95 MeV/u $\rm^{12}C$ beam on thick PMMA targets~\cite{BRAUNN20112676}. They performed another experiment on thin targets to study C-C, C-H, C-O, C-Al and C-$\rm{^{nat}Ti}$ reactions at 95 MeV/u in 2011~\cite{02}. In the same year, the FIRST collaboration performed an experiment of 400 MeV/u $\rm^{12}C$ beam on carbon target at GSI~\cite{06}. And M De Napoli et al.~\cite{07} and T. Ogawa et al.~\cite{08} also performed differernt experiments of $\rm^{12}C$ fregmentation reactions in order to test the ability of nuclear models to reproduce the fragments production.

Our team has been engaged in experimental research in heavy ion nuclear physics\cite{GE201434,CHEN201587,ZHANG2015,Ju2015}, to learn more about the fragmentation reaction of $\rm^{12}C$~, we have carried out an experiment of 80.5 MeV/u $\rm^{12}C$ beam borbarding on C, W, Cu, Au and Pb targets. The energy spectrums and cross-sections of light particles have been measured at $30^\circ$, $60^\circ$and $120^\circ$ away from beam. The experimental setup will be presented in detail in section~\ref{s2}.

\begin{figure}[!htbp]
	\centering
 	\includegraphics[width=0.8\textwidth]{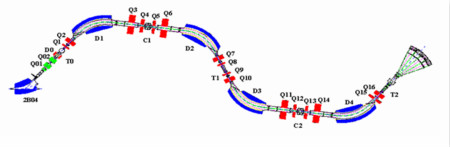}
	\caption{The schematic diagram of RIBLL.}
	\label{RIBLL}
\end{figure}
\section{Experimental setup}\label{s2}

The experiment was performed at The Radioactive Ion Beam Line in Lanzhou (RIBLL) at  the Heavy Ion Research Facility of Lanzhou (HIRFL)~\cite{SUN2003496}. Fig.~\ref{RIBLL} shows the schematic view of RIBLL.

 We have mearsured the double differential cross sections of an 80.5 MeV/u $\rm^{12}C$ beam bombarding on five different thin targets at three angles: $30^{\circ}$, $60^{\circ}$ and $120^{\circ}$. The thickness and area density ($\rm \rho ~\times$ length) of Pb, W, Cu, Au, C targets are listed in Table \ref{target}. 
 
\begin{figure}[htbp]
	\centering
	\subfigure[The target chamber from the top view.]{
 	\includegraphics[width=0.47\textwidth]{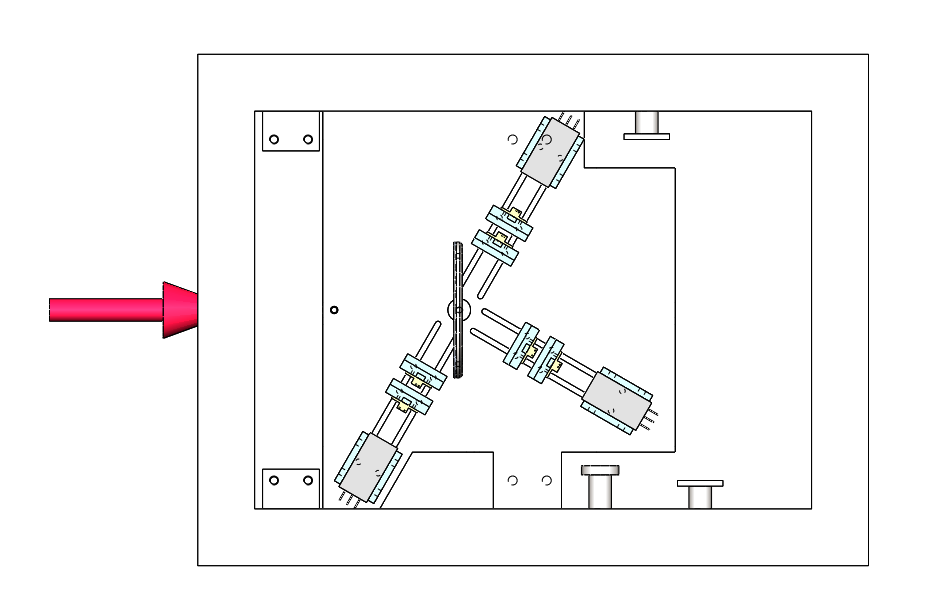}
  	}
 	\subfigure[The three dimensional schematic diagram of the target chamber.]{
 	\includegraphics[width=0.47\textwidth]{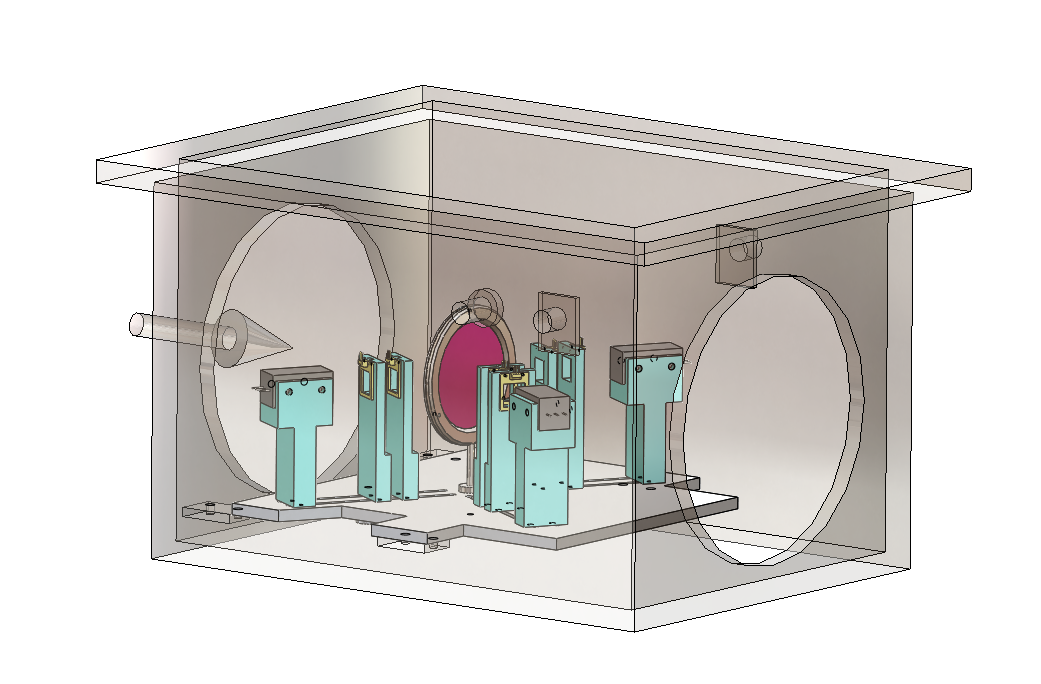}
 	}
	\caption{The schematic diagram of experiment layout.}
	\label{layout}
\end{figure}

 The experiment was carried out in T2 target chamber. The lay out of this experiment is shown in Fig.~\ref{layout}.   There is a plastic scintillator detector as beam monitor placed before the target chamber. It monitors the beam state by recording the number of particles. The beam intensity of $\rm{^{12}C}$ beam during the experiment was about $10^6$ ions/s.

\begin{table}[htbp]
	\centering 
	\caption{Parameters of targets.} 
	\label{target}
	\begin{tabular}{llllll} 
		\toprule
		Target &Pb&W&Cu&Au&C\\
		\midrule
		Thickness (mm)&0.15&0.15&0.05&0.1&1\\
		Area density ($\rm mg/cm^2$)&~170.2&290.3&44.8&193.2&185\\
		\bottomrule 
	\end{tabular} 
\end{table}

  The beam hit on the center of the target, which was fixed on the chamber. Since the targets were very thin, most of the fregments could penetrate them. 
These fregments were detected by three sets of same $\Delta E$-$\Delta E$-$E$ telescope detectors, which were fixed at  $30^{\circ}$, $60^{\circ}$ and $120^{\circ}$ with respective to the beam line, as shown in Fig.~\ref{layout}. The parameters of these telescope detectors are listed in Table~\ref{telescope}. Each set of the telescope detector consists of two Si detectors and one scintillator detector. The two Si detectors are 300 $\rm \mu m$ and 1000 $\rm \mu m$, followed by a CsI detector with a thickness of 5 cm. The distance between the center of the target and the first Si detector is 10 cm. 

\begin{table}[htbp]
	\centering 
	\caption{Parameters of one set of telescope detectors.} 
	\label{telescope}
	\begin{tabular}{llll} 
		\toprule
		~ &  Detector 1  & Detector 2   & Detector 3   \\
		\midrule
		Type&Si&Si&CsI   \\
		Thickness&300 $\rm \mu m$   &1000 $\rm \mu m$ &5 cm \\
		Distance from the target &10 cm&11.5 cm&13 cm\\
		Size &\multicolumn{3}{c}{25.4 mm $\times$ 25.4 mm}\\
		
		$\Omega$&\multicolumn{3}{c}{48.9 msr}\\
		
		\bottomrule 
	\end{tabular} 
\end{table}

The particles are measured when they have just enough energy to pass through the first Sillion detector and reach the second stage of detector. So the distence between the target and the second Stage of the detector, which is 11.5 cm, is used to calcute the acceptance of the telescope detectors. Since the size of the detectors is 25.4 mm $\times$ 25.4 mm,  the acceptance $\Omega$ is culculated to be 48.9 msr. 

\section{Data processing} \label{s3}
\subsection{Particle identification} \label{s3.1}
\begin{figure}[hbtp]
	\centering
	\subfigure[]{
 	\includegraphics[width=0.47\textwidth]{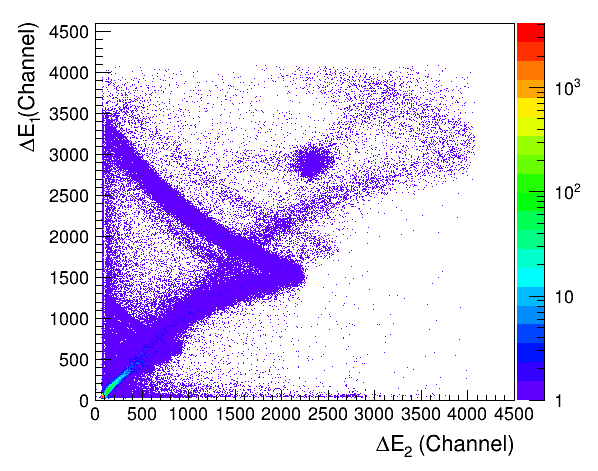}
 	\label{12}
 	}
 	\subfigure[]{
 	\includegraphics[width=0.47\textwidth]{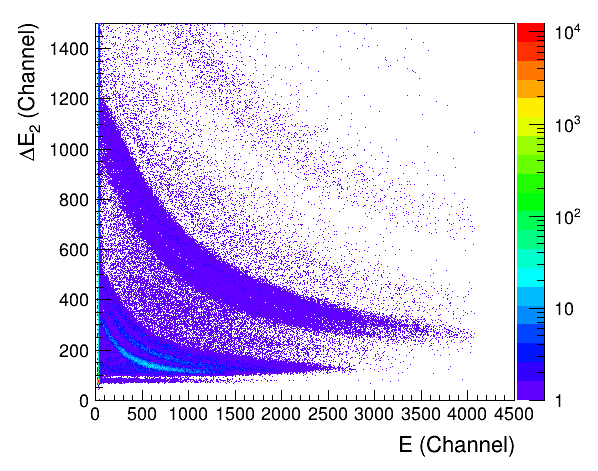}
	\label{23} 	
 	}
	\caption{The first picture is the particle identification map between the first stage and the second stage of the telescope.  The second picture is the particle identification map between the second stage and the third stage of the telescope. Both of them are the identfication map of target-Pb at 30$^\circ$.}
	\label{identification map}
\end{figure}

Particles with the same incident energy and different mass and charge lose different energies in the same material. Therefore, the particles can be identified by the telescope detectors because the deposite energy is related to their mass and charge.
 Fig.~\ref{identification map} represents the particle identification map between the first sillicon detector and the second sillicon detector and the map between the second sillicon detector and the CsI detector. It is clear that there are several separate belts in the maps. It is explicit to distinguish the particles - p, d, t, $\rm ^3He$ and $\rm ^4He$, as shown in Fig.~\ref{identification map}. The left-hand figure shows the particles punch through the first stage of the telescope, which is a sillicon detector of 300 $\mu$m. And the right-hand side figure only presents the particles which have enough energy to punch through 1300 $\mu$m sillion.

\subsection{Energy calibration} \label{s3.2}
\begin{figure}[htbp]
	\centering
	\subfigure[]{
 	\includegraphics[width=0.47\textwidth]{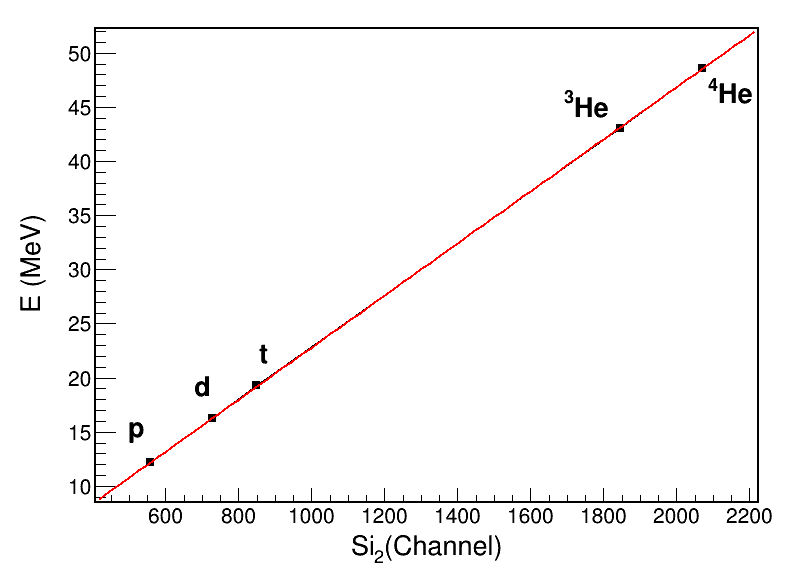}
 	\label{calib1}
 	}
 	\subfigure[]{
 	\includegraphics[width=0.47\textwidth]{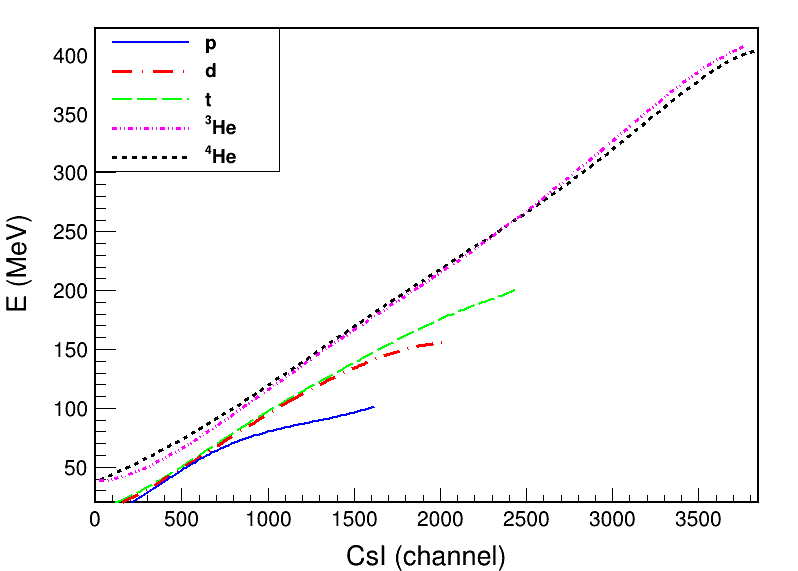}
 	\label{calib2}
 	}
	\caption{The calibration of the (a) second  and the (b) third stage of the telescope.}
	\label{calib}
\end{figure}

Energy calibration is the most important step in data processing, which is related to the double differential cross-section of the particles. Fig.~\ref{12} is used to calibrate sillion detectors. Since the horizontal axis and the vertical axis are the channel of the second and the first stage of the axis, the back bending of each belt corresponds exactly to the particles which have just enough energy to punch through the second stage of the telescope detector. The rate of energy loss of each particle and each energy is available through the SRIM software~\cite{1980iv}. With the corresponding of the punch through energies obtained by SRIM and the punch through channels obtained by the identification map, the calibration curves of the sillion detectors are obtained. One sillicon detector is calibrated only once, because its linear relationship is independent of the charge and mass of the particles. As shown in Fig.~\ref{calib1}, the calibration curve of 1000$\mu$m-thick sillicon detector is determined by five particles: p, d, t, $\rm ^3 He$ and $\rm ^4He$. 

The calibration of CsI detectors is much more complex compared with sillicon detectors. Since the ratio at which a scintillator converts the absorbed radiation energy into photons varies for different particles, each fragment needs to calibrate for CsI once, as shown in Fig.~\ref{calib2}. First, considering the linear relationship between the channel and the energy of the thick sillion detector, the verticle axis of Fig.~\ref{23} can be replaced as the energy of the thick sillicon detector. Therefore, we can get the relationship between the channel of the CsI detector and energy deposite in the thick sillicon detector. Second, thanks to the SRIM software, it is possible to get the energy deposite in the CsI detector knowing the energy deposite in the thick sillicon detector. Finally, corresponding to step one and step two, we can get the correspondence between the channel and the energy of the CsI detector.

\section{Experimental results} \label{s4}
\subsection{Energy spectrum at 30$^\circ$}
\begin{figure}[htbp]
	\centering
	\subfigure[]{
 	\includegraphics[width=0.47\textwidth]{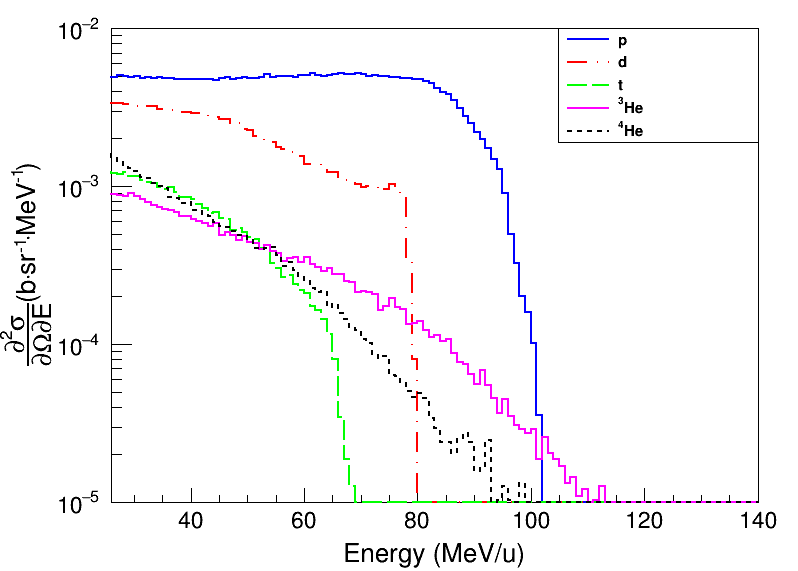}
 	\label{a-C}
 	}
 	\subfigure[]{
 	\includegraphics[width=0.47\textwidth]{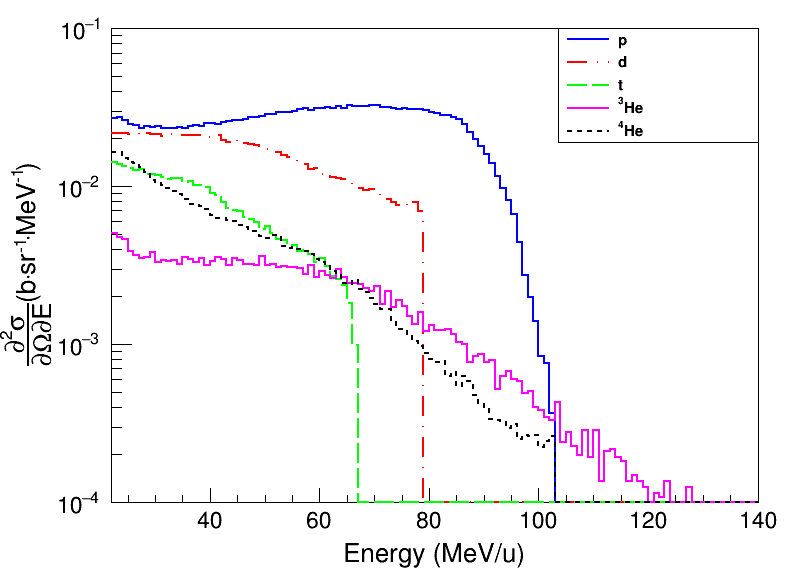}
 	\label{a-W}
 	}
 	 \subfigure[]{
 	\includegraphics[width=0.47\textwidth]{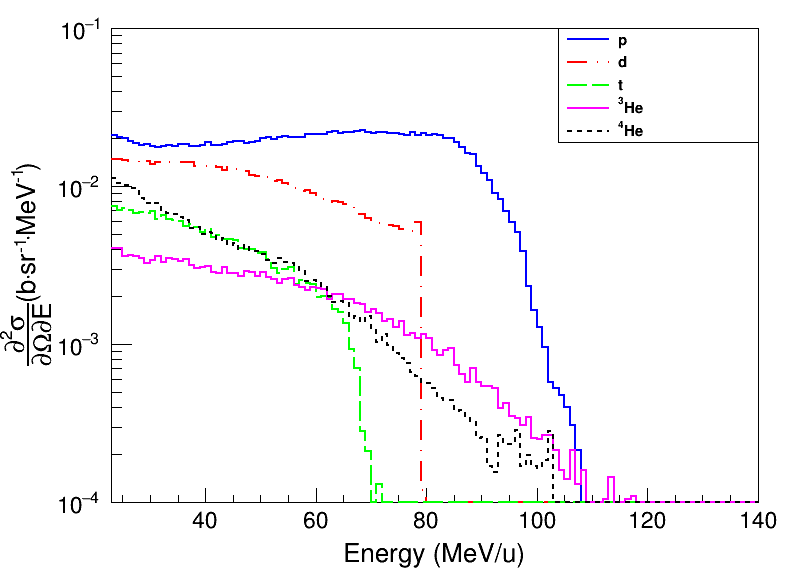}
 	\label{a-Cu}
 	}
 	 \subfigure[]{
 	\includegraphics[width=0.47\textwidth]{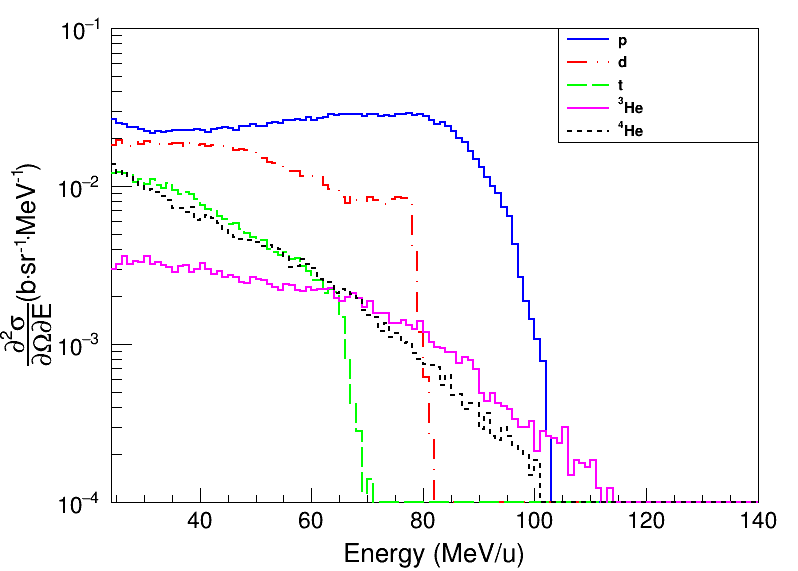}
 	\label{a-Au}
 	}
 	 \subfigure[]{
 	\includegraphics[width=0.47\textwidth]{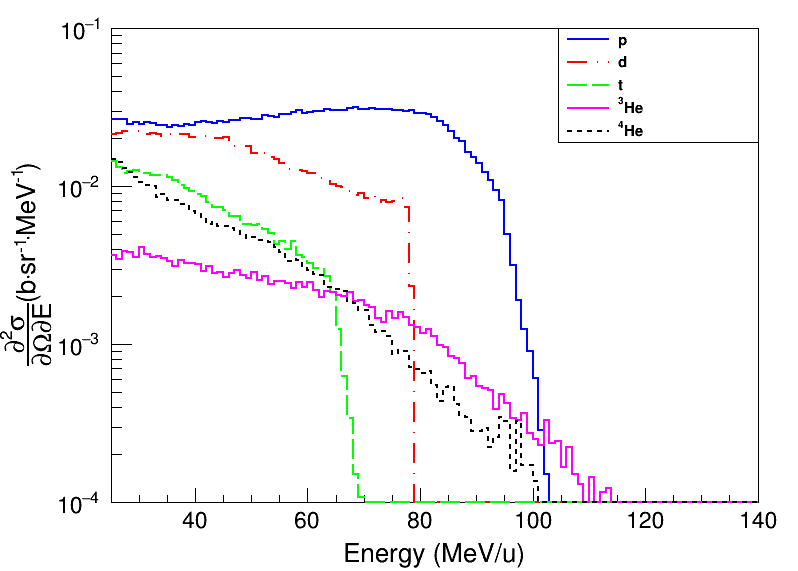}
 	\label{a-Pb}
 	}
	\caption{Energy spectrums for different particles emitted at 30$^\circ$ for (a) target-C, (b) target-W, (c) target-Cu, (d) target-Au and (e) target-Pb.}
	\label{a}
\end{figure}

After the particle identification and energy calibration, it is easy to achieve the double differential cross-sections of each fragment varies with energy at differents angles.

Fig.~\ref{a} shows the energy spectrums for different particles emitted at 30$^\circ$ for five different targets. The five lines in each figure correspond to five kinds of fragments emitted at 30$^\circ$. Compare the figures of different targets, it is clear that the same type of fragment has similar shapes of energy spectrum, regardless of targets. Besides, for isotopes of Z = 1, the double differential cross-sections decrease as A increases. If we pay attention to the details, the double differential cross-sections of protons basically remain unchanged until 80 MeV/u, which is close to our incident energy. And the energy spectrums of deutrons keep flat before 45 MeV and then go down. It is also worth noting that the energy spectrums of $\rm ^3He$ and $\rm ^4He$ have an intersection at about 60 MeV/u.
\begin{figure}[htbp]
	\centering
	\subfigure[]{
 	\includegraphics[width=0.47\textwidth]{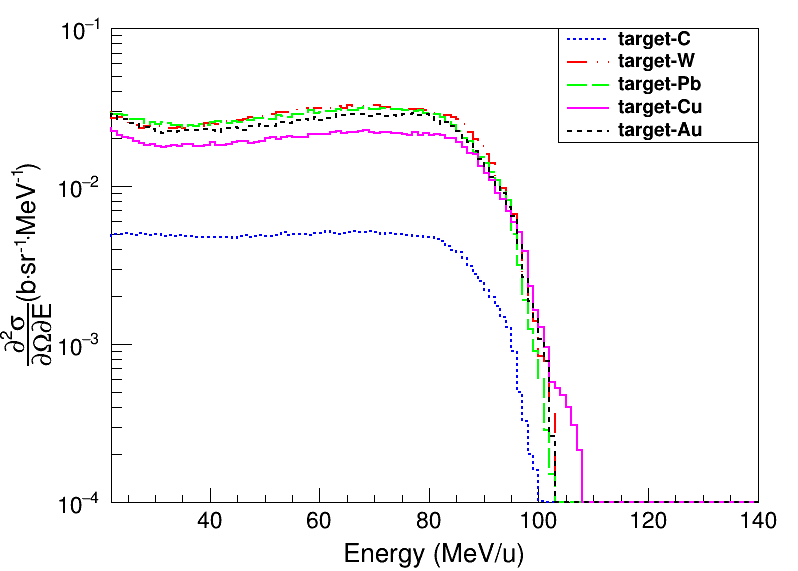}
 	\label{b-p}
 	}
 	\subfigure[]{
 	\includegraphics[width=0.47\textwidth]{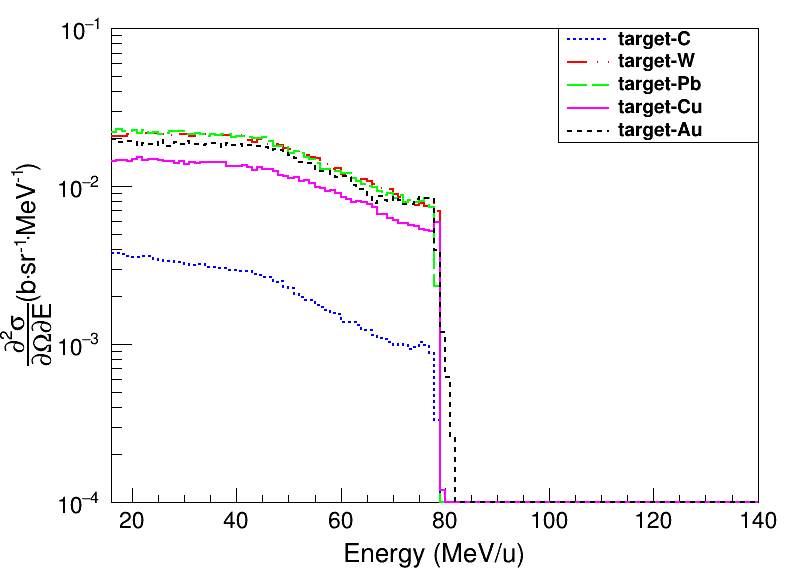}
 	\label{b-d}
 	}
 	 \subfigure[]{
 	\includegraphics[width=0.47\textwidth]{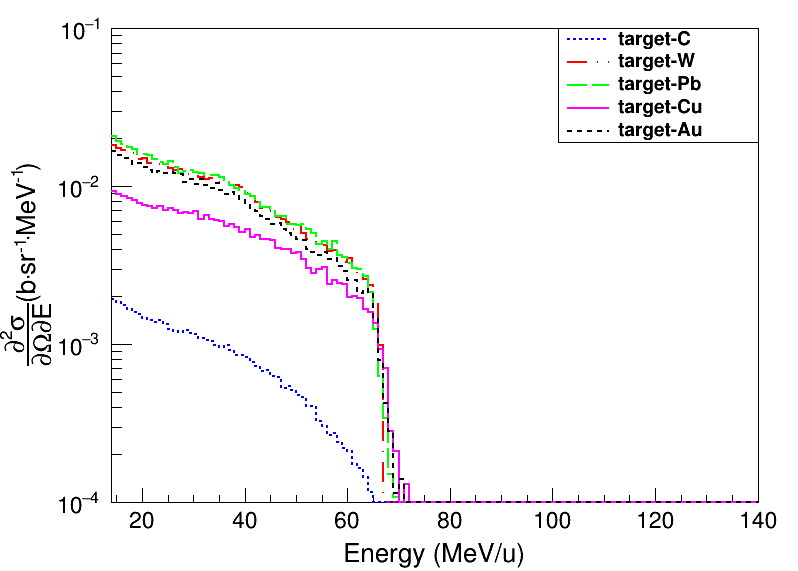}
 	\label{b-t}
 	}
 	 \subfigure[]{
 	\includegraphics[width=0.47\textwidth]{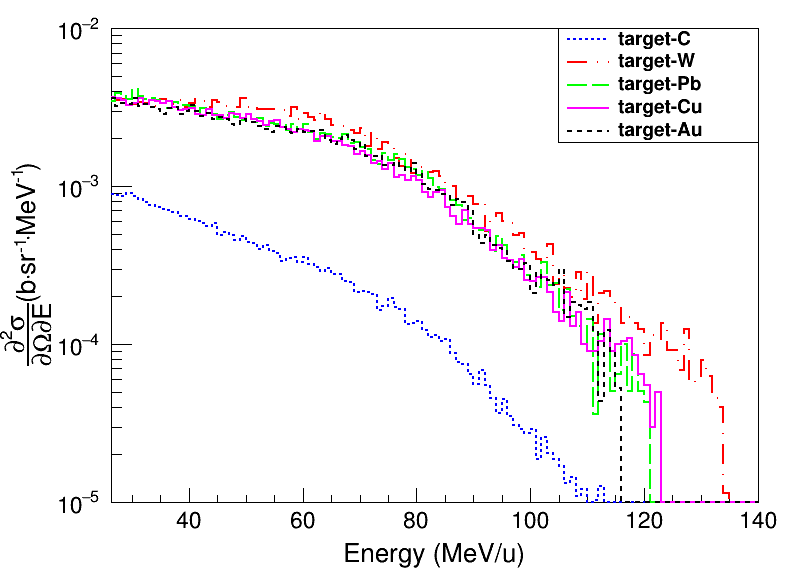}
 	\label{b-3He}
 	}
 	 \subfigure[]{
 	\includegraphics[width=0.47\textwidth]{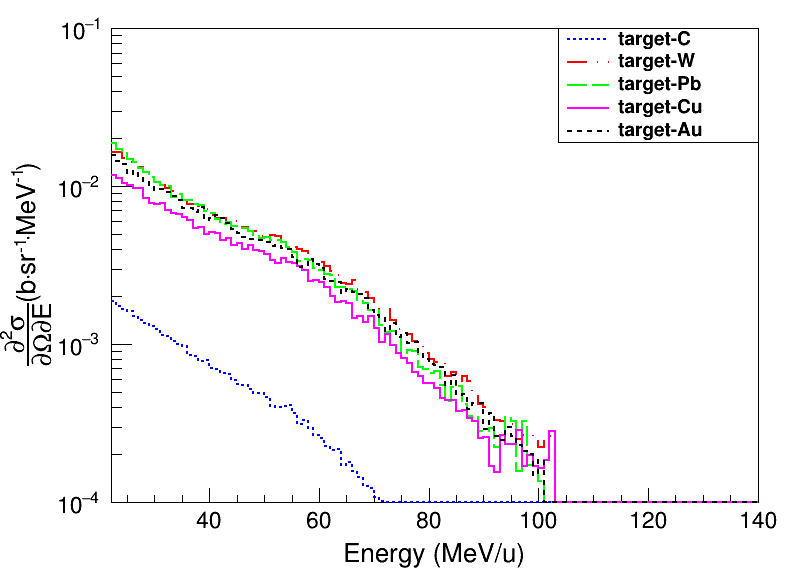}
 	\label{b-4He}
 	}
	\caption{Energy spectrums for (a) p, (b) d, (c) t, (d) $\rm ^3He$ and (e) $\rm ^4He$ emitted at 30$^\circ$ for different targets.}
	\label{b}
\end{figure}

For a more intuitive observation, each picture in Fig.~\ref{b} combines the enrgy spectrum of the same fragments at different targets at 30$^\circ$. It is obvious that the same fragments have similar energy spectrum distribution for different targets, as proved in Fig.~\ref{a}. In addition, take Fig.~\ref{b-p} as an example, the double differential cross-section of protons for C target at 30$^\circ$ is smaller than other heavier targets for less than an order of mangnitude. And that for Cu target is also a little bit smaller than  W, Pb and Au, which may lead to a conclusion that heavier targets have larger double differential cross-sections of fragments. However, the energy spectrums for W, Pb and Au targets are indistinguishable, which makes this conclusion more like a tendency.
In Fig.~\ref{b}, we can also observe that the energy spectrums of protons and deutrons have a flat area at low energies and then decrease. However, this phenomenon has not been seen in other fragments.

\subsection{Energy spectrum at different angles}
\begin{figure}[htbp]
	\centering
	\subfigure[]{
 	\includegraphics[width=0.47\textwidth]{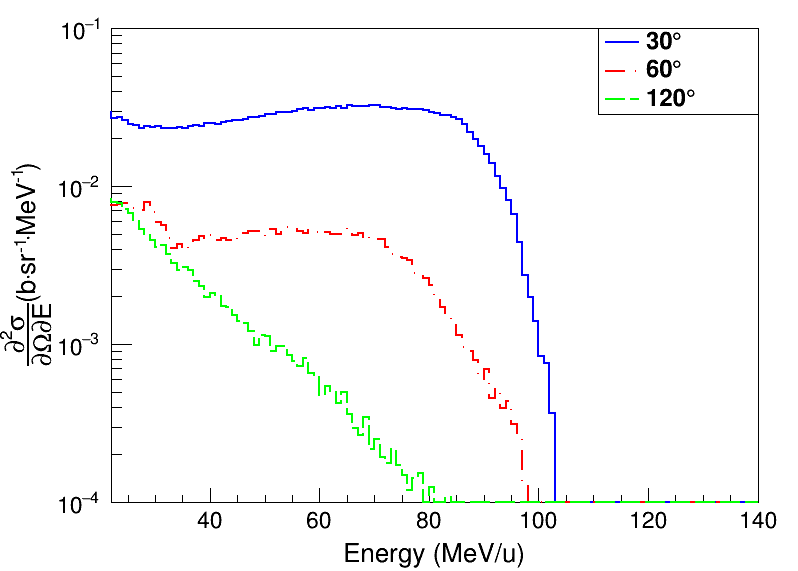}
 	\label{c-p}
 	}
 	\subfigure[]{
 	\includegraphics[width=0.47\textwidth]{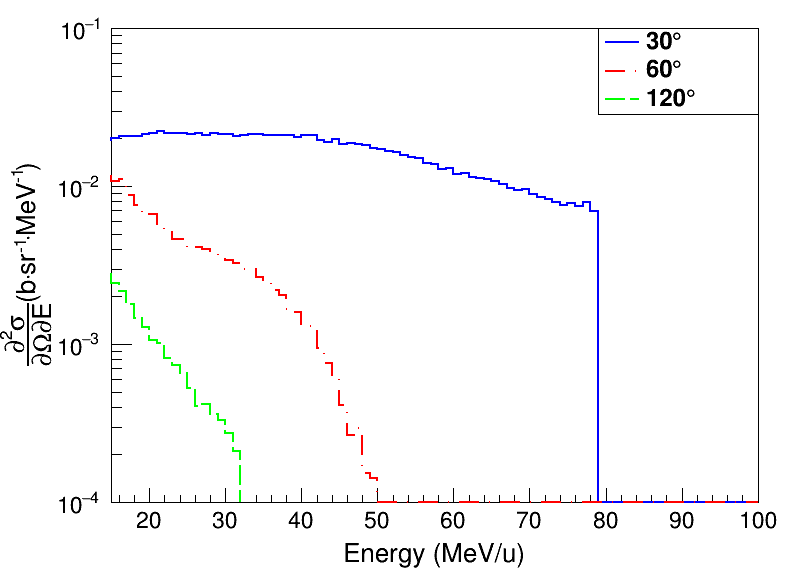}
 	\label{c-d}
 	}
 	 \subfigure[]{
 	\includegraphics[width=0.47\textwidth]{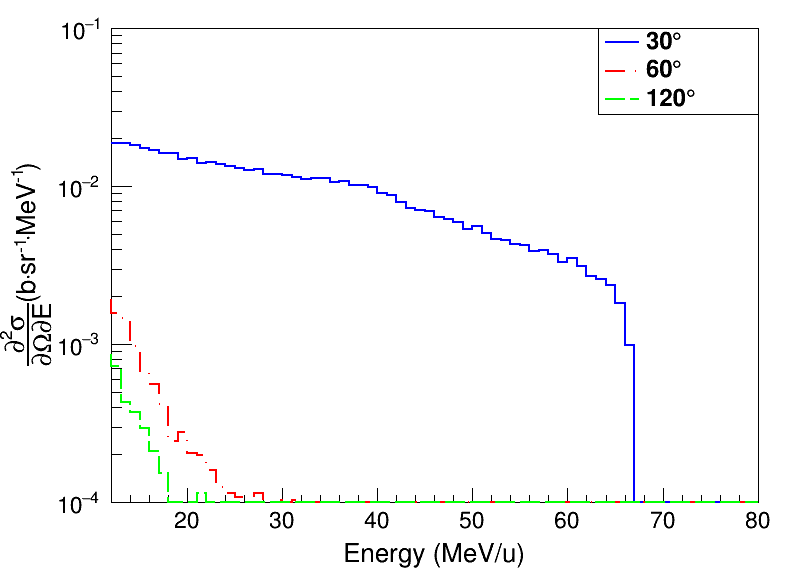}
 	\label{c-t}
 	}
 	 \subfigure[]{
 	\includegraphics[width=0.47\textwidth]{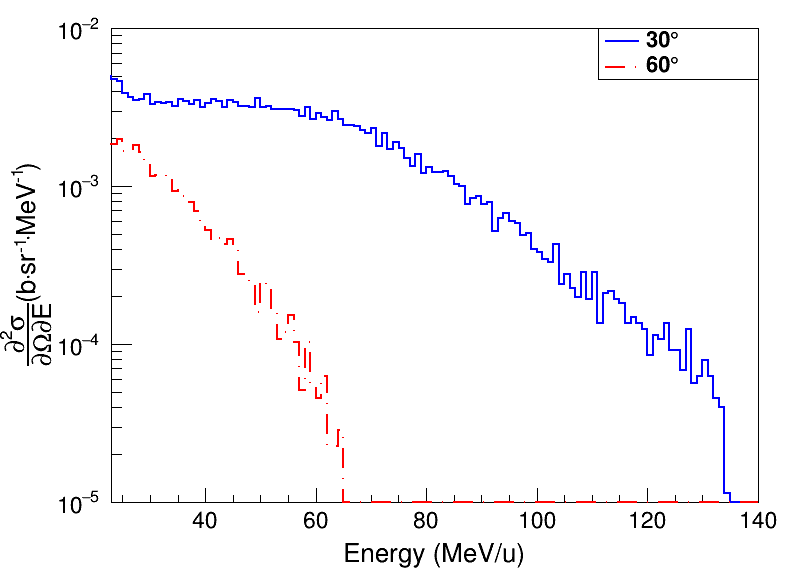}
 	\label{c-3He}
 	}
 	 \subfigure[]{
 	\includegraphics[width=0.47\textwidth]{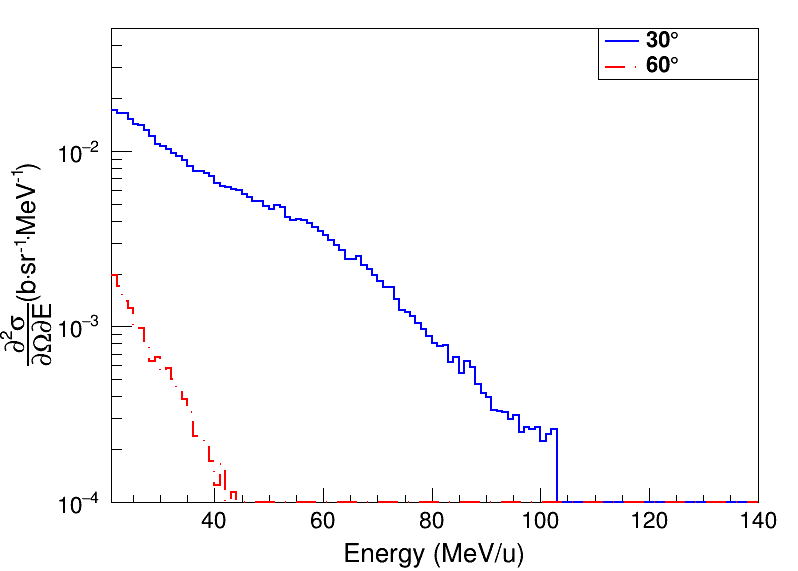}
 	\label{c-4He}
 	}
	\caption{Energy spectrums for (a) p, (b) d, (c) t, (d) $\rm ^3He$ and (e) $\rm ^4He$ emitted at $30^\circ$, $60^\circ$ and $120^\circ$ for tungsten target.} 
	\label{c}
\end{figure}
To describe the spatial distribution of the fragments, Fig.~\ref{c} exhibites the energy spectrums for different particles emitted at $30^\circ$, $60^\circ$ and $120^\circ$ for tungsten target$\footnote{The data of Cu, Au and Pb targets at 60$^\circ$ and $120^\circ$ was not obtained due to detector malfunction.}$. It is obvious that as the emitted angle become larger, the fragments of high energy are much less and the energy spectrum looks more like a straight line when the vertical axis is logarithmic. For isotopes of Z = 2, the emitted fragments at 120$^\circ$ are too few to detect. Besides, there is a more significant difference of double differential cross-sections between the three angles for heavier fragments.

\subsection{Cross-section}\label{s4.3}

\begin{figure}[htbp]
	\centering
	\includegraphics[width=0.8\textwidth]{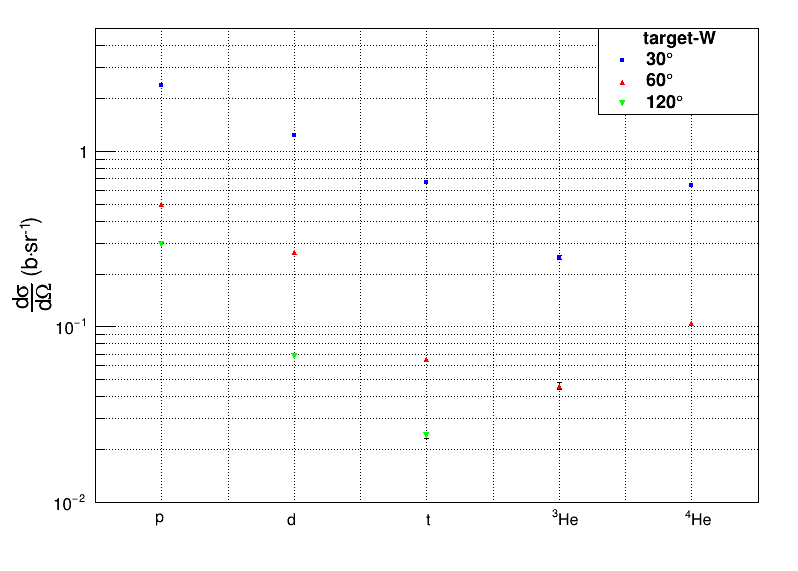}
	\caption{The cross-sections of different particles emitted at $30^\circ$, $60^\circ$ and $120^\circ$ for tungsten target (only statistical errors are presented).}	\label{d}
\end{figure}

The cross-sections have been obtained by integrating the double differential cross-sections of energy. Fig.~\ref{d} presents the cross-sections of different particles emitted at  $30^\circ$, $60^\circ$ and $120^\circ$ for tungsten target. This figure can be interpreted in two ways. For a certain emitted angle, the cross-sections of p, d and t decrease exponentially. Second, for a certain kind of fragment, the cross-sections decrease as the emitted angles increase, moreover, this kind of decrease seems to follow some kind of pattern. take $\rm ^4{He}$ for example, the cross-section difference of $30^\circ$ and $60^\circ$ is close to the cross-section difference of $\rm ^3He$.

\section{Summary}
An experiment of 80.5 MeV/u $\rm ^{12}C$ beam bombarding on C, W, Cu, Au, Pb targets at different angles has been performed at RIBLL. The emitted particles of $Z \leq 2$ were detected at $30^\circ$, $60^\circ$ and $120^\circ$ relative to the incident beam. The methods of particle identification and energy calibration are introduced in section~\ref{s3}, and the energy spectrums and cross-sections are presented in section~\ref{s4}. It is apparently that energy spectrums of the same fragments, the same exit angles and different targets have similar shape. The heavy targets have larger double differential cross-sections. At the same time, the double differential cross-sections of target-C are the smallest when all other variables are the same. The cross-sections of the emitted fragments decrease with the increasing emitted angles. Besides, there may be a certain rule in it.

\section{Acknowledgement}
This work was supported by the National Natural Science Foundation
of China (No. 11575267, 11775284, 12075296, U1832205, 11875298 and 11605258)

\newpage

\bibliography{mybibfile.bib}

\end{document}